\documentclass[a4paper]{article}

\usepackage{INTERSPEECH2022}
\usepackage{amsmath,graphicx}
\usepackage{multirow}
\usepackage{booktabs}
\usepackage{algorithm}  
\usepackage{algpseudocode}  
\usepackage{amsthm,amsmath,amssymb}
\usepackage{mathrsfs}
\usepackage{color}
\usepackage{xcolor}
\usepackage{bm}

\title{Towards Green ASR: Lossless 4-bit Quantization of a Hybrid \\
TDNN System on the 300-hr Switchboard Corpus}
\name{Junhao Xu, Shoukang Hu,  Xunying Liu, Helen Meng}
\address{
  The Chinese University of Hong Kong, Hong Kong SAR, China}
\email{jhxu, skhu, xyliu, hmmeng@se.cuhk.edu.hk}

\begin{document}

\maketitle
\begin{abstract}
  State-of-the-art time automatic speech recognition (ASR) systems are becoming increasingly complex and expensive for practical applications. This paper presents the development of a high performance and low-footprint 4-bit quantized LF-MMI trained factored time delay neural networks (TDNNs) based ASR system on the 300-hr Switchboard corpus. A key feature of the overall system design is to account for the fine-grained, varying performance sensitivity at different model components to quantization errors. To this end, a set of neural architectural compression and mixed precision quantization approaches were used to facilitate hidden layer level auto-configuration of optimal factored TDNN weight matrix subspace dimensionality and quantization bit-widths. The proposed techniques were also used to produce 2-bit mixed precision quantized Transformer language models. Experiments conducted on the Switchboard data suggest that the proposed neural architectural compression and mixed precision quantization techniques consistently outperform the uniform precision quantised baseline systems of comparable bit-widths in terms of word error rate (WER). An overall “lossless” compression ratio of 13.6 was obtained over the baseline full precision system including both the TDNN and Transformer  components while incurring no statistically significant WER increase. 
\end{abstract}

\section{Introduction}

State-of-the-art automatic speech recognition (ASR) systems represented by both the traditional hybrid HMM-DNN architectures~\cite{PoveySemi2018, abdelapplying2012, abdelexplore2013, gravesspeech2013, amberkarspeech2018, peddintiTDNN2015, Poveypure2016, hochreiterlstm1997, saklstm2014} and the recently emerging all neural end-to-end (E2E) modelling paradigm~\cite{chanlas2016, gravesctc2006,vaswani2017,dongspeech2018, gulanticonformer2020,huneural2021} are becoming increasingly complex and expensive for practical applications. For example, the best E2E encoder-decoder ASR system presented in~\cite{tuske2020single} contained up to 280.1 million model parameters. This not only lead to a large increase in their computational cost when operating on the cloud, but also creates difficulty when deploying them on edge devices to enhance privacy and reduce latency. 

An ultimate goal for many data and resource intensive deep learning based AI applications, including ASR systems, is to derive ``lossless" model compression approaches that allow high performance and low-footprint speech recognition systems to be constructed while incurring minimum performance degradation. To this end, one efficient solution is to use low-bit deep neural network (DNN) quantization techniques~\cite{KYu-2020,YQian-2019,CLeng-2018,RMa-2019}, which has drawn increasing interest in the machine learning and speech technology community in recent years. By replacing floating point weights with low precision values, the resulting quantization methods can significantly reduce the model size and inference time without modifying the model architectures. Traditional DNN quantization approaches~\cite{Courbariaux-2015, Jun-2020, fasoli21_interspeech} are predominantly based on uniform precision, where a manually defined identical bit-width is applied to all weight parameters. This fails to account for the varying performance sensitivity at different parts of the system to quantization errors.

This paper presents the development of a high performance and ultra-compact 4-bit quantized LF-MMI trained factored time delay neural networks (TDNNs) based ASR system on the 300-hr Switchboard corpus. A key feature of the overall system design is to account for the fine-grained, varying performance sensitivity at different model components to compression and quantization errors. This allows the trade-off between model compression ratio and accuracy performance target. The overall system development contained two stages. In first stage, DARTS~\cite{liudarts2019} based neural architectural compression approaches tailor designed for TDNNs~\cite{huneural2021}
are utilized first to minimize the model structural redundancy in full precision models first to facilitate hidden layer level auto-configuration of optimal factored TDNN weight matrix subspace dimensionality. This severs as a streamlined starting point for the following mixed precision quantization stage, where minimum KL divergence, log-likelihood curvature or mixed precision architectural search based approaches~\cite{xu2021mixed,Jun-2021, xu2022mixed}
are adopted to learn the optimal TDNN layer level quantization bit-widths.  To overcome the difficulty in using gradient descent methods to directly estimate discrete quantized TDNN model parameters, alternating direction methods of multipliers (ADMM)~\cite{Jun-2020} and quantization aware training (QAT)~\cite{nguyen2020quantization} are also used in the pre-training and fine-tuning stages before and after the locally varying precisions are determined. The proposed techniques were also used to produce layer level 2-bit mixed precision quantized Transformer language models~\cite{xu2021mixed, Jun-2021}.

Experiments conducted on the 300-hr benchmark Switchboard data suggest that the proposed neural architectural compression and mixed precision quantization techniques consistently outperform the uniform precision quantised baseline TDNN systems of comparable bit-widths in terms of word error rate (WER) reductions up to 1.5\% absolute. An overall ``lossless" compression ratio of 13.6 was obtained over the TDNN and Transformer models while incurring no statistically significant WER increase over the baseline 32-bit full precision system. In addition, if adopting a more generous 8-bit mixed precision based TDNN quantization, corresponding to an overall system compression ratio of 9.3, a marginal WER reduction was obtained over the full precision system.

The main contributions of this paper are summarized as follows. First, this paper presents the first application of neural architectural compression and mixed precision quantization techniques to state-of-the-art LF-MMI trained hybrid TDNN and Transformer LM based ASR systems that already benefit from a compact model structure design using low-rank weight matrix factorization~\cite{PoveySemi2018}. Latest research~\cite{david2022end2end} combines neural architecture search and quantization to key word spotting tasks, but manual defined precision setting is applied to weight and activation parameters. Similar approaches using uniform or manually configured quantization precision were previously investigated only in the context of hybrid BLSTM and RNN transducer based systems~\cite{fasoli21_interspeech, nguyen2020quantization}. Second, a largest “lossless” system compression ratio of 13.6 was obtained over the full precision 32-bit baseline without statistically significant WER increase on the benchmark 300-hr Switchboard data. This is in contrast to prior researchers using low-bit quantization approaches on the same task where statistically significant accuracy performance degradation was reported~\cite{fasoli21_interspeech}. 

The rest of this paper is organized as follows. TDNN system with low-rank weight factorization is reviewed in Section 2. Neural architecture search based TDNN model compression approaches are presented in Section 3. Uniform precision neural network quantization methods are reviewed in Section 4. Section 5 presents three mixed precision quantization methods. Experiments and results are shown in Section 6. Finally, conclusions and future work are discussed in Section 7. 

\vspace{-0.5em}
\section{Time Delay Neural Network}
\vspace{-0.5em}
Time delay neural networks (TDNNs)~\cite{waibel1989phoneme,waibel1988consonant} based hybrid HMM-DNN acoustic models in recent years defined state-of-the-art speech recognition performance over a wide range of tasks. In particular, the recently proposed lattice-free MMI trained factored TDNN systems~\cite{gulanticonformer2020} benefits from a compact model structure featuring low-rank weight matrix factorization while remain highly competitive performance wise against end-to-end approaches to date.

TDNNs can be considered as a special form of one-dimensional convolutional neural networks (CNNs)~\cite{lecuncnn1995}  when parameters are tied across different time steps. The bottom layers of TDNNs are designed to learn a narrower temporal context span, while the higher layers to learn wider, longer range temporal contexts. To further reduce the risk of overfitting to limited training data and the number of parameters, a factored TDNN (TDNN-F) model structure was proposed in~\cite{PoveySemi2018}, which compresses the weight matrix by using semi-orthogonal low-rank matrix factorization. In the TDNN-F model, the hidden layer specific bottleneck projection dimensionality settings present the group of hyper-parameter that needs to be determined in this paper using neural architecture compression prior to layer level low precision quantization is applied. 

\vspace{-1.5ex}
\subsection{Neural Architecture Search}

\begin{table*}[h]
  \caption{\emph{Performance (WER\%, $\#$parameters) comparison of TDNN-F models configured with projection dimensionality of the Kaldi recipe baseline system, and the pipelined Gumbel-softmax DARTS (PipeGumbel) system. $\eta$ is the penalty factor. The dimensionality choices: \{25,50,80,100,120,160,200,240\}. * denotes a statistically significant WER difference over the baseline system (sys. 1). }}
  \vspace{-0.7em}
  \label{tab:swbd-trans}
  \centering
\resizebox{150mm}{8mm}{
  \begin{tabular}{c|c|c|c|cc|ccc|cc|c|c} 
      \toprule
      \multirow{2}{*}{\textbf{ID}}&  \multirow{2}{*}{\textbf{Method}}& \multirow{2}{*}{\textbf{$\eta$}} &\multirow{2}{*}{\textbf{Bottleneck Dim Index}} &  \multicolumn{2}{c|}{\textbf{eval2000}}&
      \multicolumn{3}{c|}{\textbf{rt02}}&
      \multicolumn{2}{c|}{\textbf{rt03}}& \textbf{WER(\%)} & 
  \multirow{2}{*}{\textbf{\#param}} \\  
      & & & & swbd & callhm & swbd1 & swbd2 & swbd3 & fsh. & swbd & avg.& \\ \midrule
      1 & Baseline & - & 5 5 5 5 5 5 5 5 5 5 5 5 5 5 & 9.7 & 18.0 & 11.5 & 15.3 & 20.0 & 12.6 & 19.5 & 15.5 & 18.6M \\
      2 & PipeGumbel & 0.1 & 2 3 2 0 2 2 2 1 2 1 0 3 5 5 & 9.2* & 17.4* & 10.9* & 14.4* & 19.5* & 11.9* & 19.2* & 15.0* & 12.4M \\\bottomrule
      \end{tabular}
}
\vspace{-2em}
\end{table*}
In this section, we introduce the pipelined Gumbel-softmax differentiable neural architecture search (DARTS) methods adopted in this paper. The general form of DARTS architecture selection methods~\cite{liudarts2019, dongsearch2019}are introduced as follows. 
For example, the l-th neural network hidden layer output $\bm{h}^l$ can be computed as a linear combination between the architecture weights $\lambda_i^l$ and candidate architecture choices $\phi_i^l(\cdot)$ in the DARTS super-network:
\vspace{-1em}
\begin{equation}
    \bm{h}^l = \sum^{N^l-1}_{i=0}\lambda^l_i\phi^l_i(\bm{W}_i^l\bm{h}^{l-1})
\vspace{-1em}
\end{equation}
where $\lambda_i^l$ is the architecture weight for the i-th candidate
choice in the l-th layer, $N_l$ is the total number of choices
in this layer. The precise form of neural architectures being
considered at this layer is determined by the linear transformation $\bm{W}^l_i$ and activation function $\phi$(·) used by each candidate system. For example, when selecting the dimensionality of the bottleneck linear projection at each hidden layer, the linear transformation $\bm{W}_i^l = \Tilde{\bm{W}}_i^l\Tilde{\bm{W}}_i^{lT}$ is a 
decomposed matrix of the $i^{th}$ projection dimensionality, while $\phi(\cdot)$ is also an identity matrix. 
\vspace{-1em}
\subsection{Gumbel-softmax DARTS}
\vspace{-0.5em}
For traditional DARTS methods, when similar architecture weights are obtained using a flattened Softmax function, the confusion over different candidate systems increases and search errors may occur. 
To this end, a Gumbel-Softmax distribution~\cite{dongsearch2019, xie2019snas, maddison2017the} is used to sharpen the architecture weights to produce approximately a one-hot vector. This allows the confusion between different architectures to be minimised. The architecture weights are computed as,
\begin{equation}
    \lambda_i^l = \frac{exp((log\gamma_i^l+G_i^l)/T)}{\sum_{j=0}^{N^l-1}exp((log\gamma_j^l+G_j^l)/T)}
\end{equation}

\noindent where $G_i^l = -log(-log(U_i^l))$ is the Gumbel variable, and $U_i^l$ is a uniform random variable. When the temperature parameter $T$ approaches 0, it has been shown that the Gumbel-Softmax distribution is close to a categorical distribution~\cite{maddison2017the}. Different samples of the uniform random variable $U_i^l$ lead to different values of $\lambda_i^l$ in Eq. 1. The loss function gradient w.r.t $log\gamma_k^l$ is computed as an average over J samples of the architecture weights,
\vspace{-1em}
\subsection{Pipelined DARTS}
\vspace{-0.5em}
For the above Gumbel-Softmax DARTS systems, the search algorithms may prematurely select sub-optimal architectures at an early stage. Inspired by~\cite{guosingle2020}, we decouple the update of normal DNN parameters and architecture weights into two separate stages performed in sequence. This leads to the pipelined DARTS approach. In order to prevent overfitting to the training data, a separate held-out data set taken out of the original training data is used. we randomly select 5\% of the original training set as the held-out data set. In Pipelined DARTS systems, the normal DNN parameters are updated to convergence on the training data first (not containing the separate held-out data), while randomly sampled one-hot architecture weights drawn from a uniform distribution are used. In the following stage, we fix the normal DNN parameters estimated in the first stage in the super-network and update the architecture weights using the held-out data for the Gumbel-Softmax DARTS. This produces the Pipelined Gumbel-softmax DARTS (PipeGumbel) systems.
To further adjust the trade-off between system performance and compression ratio, a penalized loss function incorporating the underlined neural network size is used, given as follows: 
\begin{equation}
  L = L_{LF-MMI} + \eta\sum_{l,i}\lambda_i^lC_i^l
  \vspace{-1em}
\end{equation}
\noindent where $L_{LF-MMI}$ is the lattice-free MMI criterion. $C_i^l$ is the complexity penalty term expressed as the model parameters of the i-th candidate considered at the l-th layer. $\eta$ is the penalty scaling factor empirically set for different tasks. 

\vspace{-0.5em}
\section{Neural Network Quantization}
\vspace{-0.5em}
For a standard $n$-bit quantization problem of neural networks, we consider a full precision weight parameter $\Theta$ and find its closest discrete approximation from the following quantization table $Q \in \{0, \pm1,\pm2, \dots, \pm (2^{n-1}-1)\}$ as
\begin{equation}
 f(\Theta) = \arg\min \limits_{Q} |\Theta - Q|
 \vspace{-0.5em}
\end{equation}
while one bit is reserved to denote the sign. With further simplification, low bit quantization, for example, binarization $\{1, -1\}$~\cite{Rastegari-2016, CLeng-2018} and ternary $\{-1, 0, 1\}$~\cite{LiF-2016}, can be produced.

When applying quantization to all weight matrices in the model, we can use a more general format in equation (5) to represent the quantization for each parameter. Let $\Theta^{(l)}_{i}$ be the $i^{th}$ parameter within any of the $l^{th}$ weight cluster, for example, all weight parameters of the same TDNN hidden layer,

\vspace{-1em}
 \begin{equation}
  \begin{split}
  \label{eq:quanMap}
  f(\Theta^{(l)}_i) = \arg\min \limits_{Q_i^{(l)}} |\Theta^{(l)}_i - Q_i^{(l)}|
\end{split}
\end{equation}
The locally shared $l^{th}$ quantization table is given by
\begin{equation}
\label{equ:31}
  {Q}_i^{(l)}\in\{0, \pm\alpha^{(l)},\dots, \pm\alpha^{(l)} (2^{n_{l}-1}-1)\}
\end{equation}
where $\alpha^{(l)}$ is a full precision scaling factor used to adjust the dynamic range of all the unquantized weights in the cluster. It is shared locally among weight parameters clusters. A special case, when the local quantization table in equation (5) is shared across all the layers, leads to the traditional uniform precision quantization approach. The only remaining factor affecting the system performance is the bit length $n_{l}$ which is also globally set to be $1, 2, 4, 8, 16$ etc. 
Efficient alternating direction methods of multipliers (ADMM)~\cite{Jun-2020} based optimization of local quantization tables can be used.

\vspace{-1.5ex}
\section{Mixed Precision Quantization}
\vspace{-0.5ex}
\label{sec:mixedquant} 
This section presents three approaches to automatically learn the optimal local precision settings for TDNN-F systems.

\vspace{-1.5ex}
\subsection{KL Divergence Based Mixed Precision Quantization}
 In order to minimize the distance between the distribution embodied by the full precision system and that of the quantized model, Kullback-Leibler (KL) divergence between full precision and quantized NNs is used to measure the resulting performance sensitivity.  
Taking a $L$-layer TDNN-F for example, for any quantization $f(\cdot)$ being applied to the full precision parameters ${\bf \Theta}$, the KL divergence based quantization sensitivity measure is computed over the input spectrum of $T$ frames: 
\vspace{-1em}
\begin{align}
\vspace{-1em}
  \Omega^{\rm{KL}} &= \sum_{i=1}^L\Omega^{\rm{KL}}_i = \sum_{i=1}^L D_{\rm{KL}}(P(\bm{\Theta_i})||P(f_{n_i}(\bm{\Theta_i})))\\
  & =\sum_{i=1}^{L} \sum_{t=1}^{T}P(\sigma(\bm{h}^{t}_{l}), {\bf \Theta}_i)\ln{\frac{P(\sigma(\bm{h}^{t}_{l}), {\bf \Theta }_i)}{P(\sigma(\bm{h}^{t}_{l}), f_{n_i}({\bf \Theta}_i))}}\notag
 \end{align}
 \vspace{-1em}
 
\noindent where ${\bf \Theta}_i$ denote the full precision parameters of the $i^{th}$ layer, and $f_{n_i}({\bf \Theta}_i)$ is $n_{i}$-bit quantized parameters given a particular local precision bit width $n_i$ and ${\bm h}^{l}_{t}$ is the TDNN-F output vector computed at layer $l$. When computing the KL metric in Eqn. (7), ${\bm h}^{l}_{t}$ is fed into a Sigmoid gate first to produce normalised, probability like outputs between 0 and 1, $P(\sigma(\bm{h}^{l}_{t}), {\bf \Theta})$.
Given a target model size constraint (e.g. average 4-bit precision), the KL metric for each precision setting of each layer is computed and minimized to select the optimal local bit-width while satisfying the constraint\footnote{Based on the performance sensitivity ranking measured by either the KL metric of Eqn. (7), or the curvature metric of Eqn. (8), the optimal local bit-widths combination that is closest to the target average quantization precision, e.g. 4-bit, while producing the minimum KL or curvature measured performance sensitivity, will be selected.}.

\vspace{-1.5ex}
\subsection{Curvature Based Mixed Precision Quantization}
The second approach minimizes the performance sensitivity to quantization by examining the local training data loss function curvature. Under mild assumptions such that the parameters of a DNN is twice differentiable and while converging to a local optimum, it is shown in ~\cite{ZDong-2019-2, ZDong-2019} that the separation performance sensitivity to quantization, when using a given precision setting, can be expressed as the squared quantization error further weighted by the parameter Hessian matrix trace. For any quantization $f(\cdot)$ being applied to the parameters $\boldsymbol{\Theta}$ of the L-layer TDNN-F acoustic model, the total performance sensitivity is given by the sum of Hessian trace weighted squared quantization error, to be minimized under a target model size constraint.
\vspace{-0.5em}
\begin{equation}
  \Omega^{\rm{Hes}} = \sum_{i=1}^L\Omega^{\rm{Hes}}_i = \sum_{i=1}^L Tr(\bm{H}_i)\cdot||f(\bm{\Theta}_i)-\bm{\Theta}_i||_2^2
\end{equation}
\vspace{-1em}

\noindent
An efficient Hessian-free approach based on the Huchinson’s Algorithm~\cite{ZDong-2019, Avron-2011} is used to approximate the Hessian trace.



\begin{table*}[h]
  \caption{\emph{Performance of the baseline full precision, uniform precision and mixed precision quantized TDNN-F systems with local precision settings automatically learned using the HES/KL/NAS methods introduced in Section 4 on Switchboard NIST Hub5'00, RT02 and RT03. RNNLM rescoring also applied. * and ${\dag}$ denote no statistically significant WER difference obtained over the two full precision systems (sys.1, sys.0) respectively. Speech processing real time factors (RTFs) in the last column.}} 
  \vspace{-0.7em}
  \label{tab:swbd-trans}
  \centering
\resizebox{170mm}{36mm}{
  \begin{tabular}{c|c|c|c|c|cc|cc|ccc|c|c|c|c} 
      \toprule
\multirow{2}{*}{\textbf{ID}}&  \textbf{quant.}&\textbf{param.} & \textbf{quant.}&\multirow{2}{*}{\textbf{\#bit}} & 
      \multicolumn{2}{c|}{\textbf{Hub5'00}}&
      \multicolumn{2}{c|}{\textbf{rt03}}&
      \multicolumn{3}{c|}{\textbf{rt02}}&\textbf{WER(\%)} &
  \textbf{model} & \textbf{comp.} & \textbf{RTF}  \\  
      &\textbf{prec.}& \textbf{estim.} &\textbf{method}& & swbd & callhm & fsh. & swbd & swbd1 & swbd2 & swbd3  & avg.& \textbf{size(MB)} & \textbf{ratio} & $*10^{-3}$ \\ \midrule
0 & \multicolumn{3}{c|}{Baseline} & 32 & 9.7 & 18.0 & 12.6 & 19.5 & 11.5 & 15.3 & 20.0 & 15.5 & 74.4 & - & 0.223 \\ \cline{1-1}
1 & \multicolumn{3}{c|}{PipeGumbel} & 32 & 9.2 & 17.4 & 11.9 & 19.2 & 10.9 & 14.4 & 19.5 & 15.0 & 49.6 & 1.5 & 0.187 \\ \cline{1-2}\cline{3-16}
2 & &  & & 1 & 16.4 & 25.0 & 18.3 & 27.2 & 17.6 & 20.9 & 27.7 & 22.3 & 1.9 & 39.2 & 0.065 \\ \cline{1-1}
3 &  & BP~\cite{Courbariaux-2015} & Offline & 2 & 15.4 & 23.9 & 17.2 & 26.5 & 16.8 & 20.1 & 26.9 & 21.7 & 3.7 & 20.1 & 0.103  \\ \cline{1-1}
4 & \emph{uniform} & \emph{(modified)} & Tuning  & 4 & 11.0 & 20.1 & 13.6 & 22.5 & 13.0 & 16.2 & 22.7 & 16.9 & 6.4 & 11.6 &  0.105 \\ \cline{1-1}
5 & prec. & &  & 8 & 10.3 & 18.7 & 12.6 & 20.4 & 11.8 & 15.2 & 20.6 & 15.8${\dag}$ & 12.6 & 5.9 & 0.097 \\ \cline{1-1}
6 & & & &  16 & 9.0 & 17.2 & 11.6 & 19.1 & 10.8 & 14.2 & 19.2 & 14.8 & 25.1 & 3.0 & 0.111 \\ \cline{1-1}\cline{3-16}
7 & &  &  & 1 & 15.0 & 23.1 & 16.2 &  25.0 & 15.5 & 18.8 & 25.1 & 19.7 & 1.9 & 39.2 & 0.057  \\ \cline{1-1}
8 & &\multirow{2}{*}{QAT~\cite{nguyen2020quantization}} & Offline & 4 & 10.6 & 19.1 & 13.0 & 21.7 & 12.4 & 15.6 & 21.5 & 16.1 & 6.4 & 11.6 & 0.090   \\\cline{1-1}
9 & & & Quant & 8 & 9.7 & 18.0 & 12.2 & 19.8 & 11.2 & 14.9 & 19.9 & 15.4${\dag}$ & 12.6 & 5.9 & 0.096 \\ \cline{1-1}
10 & & &  &  16 & 9.2 & 17.5 & 11.8 & 19.3 & 11.0 & 14.3 & 19.6 & 15.0 & 25.1 & 3.0  & 0.120  \\\hline
{\bf 11} & & BP~\cite{Courbariaux-2015} & \multirow{4}{*}{Hes} & \textbf{4} & \textbf{10.2} & \textbf{18.6} & \textbf{12.4} & \textbf{19.8} & \textbf{11.5} & \textbf{14.9} & \textbf{20.0} & \textbf{15.6}${\dag}$ & \textbf{6.3} & \textbf{11.8} & 0.073 \\ \cline{1-1}
12 & & \emph{(modified)} & & 8 & 9.5 & 17.5 & 12.1 & 19.5 & 11.1 & 14.8 & 19.6 & 15.2* & 12.6 & 5.9 & 0.077  \\ \cline{1-1}\cline{5-16}\cline{3-3}
13 & &  \multirow{2}{*}{QAT~\cite{nguyen2020quantization}} &  & 4 & 9.7 & 17.7 & 12.3 & 19.8 & 11.3 & 15.0 & 19.7 & 15.5${\dag}$ & 6.4 & 11.6 & 0.071 \\ \cline{1-1}
14 & & &  & 8 & 9.3 & 17.5 & 12.1 & 19.5 & 11.2 & 14.7 & 19.5 & 15.2* & 12.5 & 6.0 & 0.080 \\ \cline{1-1}\cline{3-16}
15 & & BP~\cite{Courbariaux-2015} & \multirow{4}{*}{KL} & 4 & 10.3 & 18.5 & 12.4 & 19.9 & 11.4 & 15.0 & 20.1 & 15.6${\dag}$ & 6.3 & 11.7 & 0.074  \\ \cline{1-1}
16 & \emph{mixed} & \emph{(modified)} & & 8 & 9.4 & 17.6 & 12.1 & 19.4 & 11.0 & 14.8 & 19.7 & 15.2* & 12.6 & 5.9 & 0.079  \\ \cline{1-1}\cline{5-16}\cline{3-3}
17 & prec. &  \multirow{2}{*}{QAT~\cite{nguyen2020quantization}} &  & 4 & 9.8 & 17.8 & 12.2 & 19.7 & 11.2 & 14.9 & 19.8 & 15.4${\dag}$ & 6.4 & 11.6 & 0.069 \\ \cline{1-1}
{\bf 18} &  & & & \textbf{8} & \textbf{9.3} & \textbf{17.6} & \textbf{12.0} & \textbf{19.4} & \textbf{11.0} & \textbf{14.6} & \textbf{19.6} & \textbf{15.1*} & \textbf{12.5} & \textbf{5.9} & 0.081 \\ \cline{1-1}\cline{3-16}
19 & & BP~\cite{Courbariaux-2015} & NAS & 4 & 10.5 & 18.8 & 12.8 & 20.6 & 11.9 & 15.5 & 21.0 & 16.0 & 6.3 & 11.8 & 0.072 \\ \cline{1-1}
20 & & \emph{(modified)} & \emph{gumbel} & 8 & 9.8 & 18.0 & 12.2 & 19.7 & 11.1 & 14.7 & 19.9 & 15.4${\dag}$ & 12.6 & 5.9 & 0.078 \\ 
\bottomrule
      \end{tabular}
 }
 \vspace{-1.5em}
\end{table*}

\vspace{-1.5ex}
\subsection{Architecture Search Based Mixed Precision Quantization}
The third solution to automatically learn the optimal local quantization precision settings is to use mixed precision based neural architecture search (NAS)~\cite{Elsken-2019, Hu-2020-CVPR} approaches. 
The super-network is constructed by first using ADMM to separately pretrain each TDNN-F layers using different uniform precision settings, before treating these uniform precision as different NAS candidate at each layer. In order to avoid the trivial selection of the longest, most generous quantization bit width, these precision selection weights learning can be further constrained by a model complexity penalty term with respect to the number of bits retained after quantization, in order to obtain a target average quantization precision, for example, 4-bit, akin to the penalized LF-MMI criterion in Eq. (3).


\begin{table}[h]
  \caption{\emph{Performance of best 4-bit and 8-bit quantized TDNN-F systems after 2-bit quantized Transformer LM rescoring.  }}
  \vspace{-0.7em}
  \label{tab:swbd-trans}
  \centering
\resizebox{82mm}{18mm}{
  \begin{tabular}{c|c|c|c|c|c|c|c} 
      \toprule
       \multirow{2}{*}{\textbf{ID}} &\multirow{2}{*}{\textbf{sys.}}& \multirow{2}{*}{\textbf{AM}}&\multirow{2}{*}{\textbf{LM}} & 
      \textbf{Hub5'00}& \textbf{rt03} & \textbf{rt02} & 
  \textbf{model}  \\  
      & & &  & swbd/chm & fsh/swbd & swb1/swb2/swb3 & \textbf{size(MB)} \\ \midrule
1 & Baseline (sys. 0, Tab 2) & 32b & 32b & 9.6/17.8  & 12.4/19.3 & 11.3/15.1/19.9 & 180.4 \\
2 & PipeGumbel (sys. 1, Tab 2) & 32b & 32b & 9.0/17.3 & 11.7/19.1 & 10.8/14.2/19.3 & 155.6  \\
3 & Baseline+LHUC & 32b & 32b & 7.7/15.1  & 9.8/16.0 & 9.4/12.2/15.8 & 180.4 \\ 
4 & PipeGumbel+LHUC & 32b & 32b & 7.5/14.4  & 9.6/15.6 & 9.0/11.8/15.6 & 155.6 \\\hline
5 & TDNN-4bit (sys. 11, Tab 2) & 4b & 2b & 10.0/18.3 & 12.2/19.5 & 11.4/14.8/19.6 & 13.3 \\
6 & TDNN-8bit (sys. 18, Tab 2) & 8b & 2b & 9.2/17.3 & 11.7/19.2 & 10.8/14.4/19.4 & 19.5 \\
7 & TDNN-4bit+LHUC & 4b & 2b & 8.5/15.6 & 10.3/16.9 & 10.0/12.8/17.1 & 13.3  \\
8 & TDNN-8bit+LHUC & 8b & 2b & 7.7/14.8 & 9.9/16.0 & 9.4/12.3/16.0 & 19.5 \\ \hline
9 & IBM-BLSTM-300HR~\cite{fasoli21_interspeech} & 4b & - & 10.4/18.3 & - & - & - \\
10 & IBM-BLSTM-2000HR~\cite{fasoli21_interspeech} & 4b & - & 8.0/14.0  & - & - & - \\
11 & IBM-RNN-T-2000HR~\cite{fasoli21_interspeech} & 32/4b & - & 8.8/17.4 & - & - &  - \\
\bottomrule
      \end{tabular}
 }
 \vspace{-2em}
\end{table}

\vspace{-0.5em}
\section{Experiments}
\vspace{-0.5ex}
The baseline full precision Kaldi recipe\footnote{All of this is in published Kaldi code at https://github.com/kaldi- asr/kaldi/tree/master/egs/swbd/s5c/run.sh and https://github.com/kaldi- asr/kaldi/tree/master/egs/swbd/s5c/local/chain/tuning/run\_tdnn\_7q.sh.} TDNN-F system (line 0, Table 1 and 2) was trained on the 300-hour Switchboard corpus featuring speed perturbation and i-Vector speaker adaptation. Model training and evaluation time measurement were conducted with one thread on a single NVIDIA Tesla V100 Volta GPU. Matched pairs sentence-segment word error (MAPSSWE) based statistical significance test was performed at a significance level $\alpha = 0.05$.
  
In the first stage of system development, the penalized Pipelined Gumbel-softmax DARTS (\emph{PipeGumbel}) approach of Section 2.3 was used to compress baseline TDNN-F system’s hidden layer projection dimensions from a uniform set recipe configuration of 200 down to as low as 25. The resulting \emph{PipeGumbel} compressed TDNN-F system (line 2, Table 1 and 2) produced significant WER reductions of 0.5\% averaged across all three test sets while reducing the number of model parameters by 33\% relative. It takes about 110 GPU hours during both architecture search and subsequent model retraining.

This system serves as the streamlined starting point of a range of quantized systems shown in Table 2 (line 2 to 20). Among these, several trends can be found. First, given the same quantization precision, for example, 4-bit, all the mixed precision quantized models introduced in Section 5, including curvature based HES (sys. 11, 13), KL (sys. 15, 17) and mixed precision NAS (sys. 19) outperform the 4-bit uniform quantized model (sys. 4). Second, quantization aware training (QAT)~\cite{nguyen2020quantization} consistently outperformed the modified back-propagation (BP)~\cite{Courbariaux-2015} \footnote{Low precision quantized parameters were first used in the forward pass to compute the error loss before full precision parameters are used in the backward pass to propagate the gradients for model update.} for both uniform or mixed precision systems. 
Third, among all the mixed precision quantization methods, the lowest WER was obtained using the 8-bit KL+QAT quantized model (sys. 18\footnote{The automatically learned TDNN-F hidden layer specific bit widths of this system are \{16,8,8,1,4,4,4,2,4,2,1,4,8,8,16\}.}), producing a compression ratio of 5.9 and 0.4\% absolute WER reduction over the baseline (sys. 0). Lastly, the largest ``lossless" compression ratio of 11.8 was obtained using the 4-bit HES+QAT quantized model (sys. 11), while incurring no statistically significant WER increase over the baseline full precision model (sys. 0).

In Table 3, the recognition outputs of the above two best performing 4-bit and 8-bit quantized systems (sys. 11, 18) were then rescored using a 2-bit quantized Transformer LM derived using a KL learned mixed precision method~\cite{xu2021mixed, Jun-2021} akin to previous results (line 5, 6 in Table 3), and further speaker adapted using learning hidden unit contribution (LHUC)~\cite{swietojanski2016learning} (line 7, 8). The performance of the comparable baseline and NAS compressed full precision systems (line 1, 2) are also shown. These serve to contrast with the recently published low-bit quantized ASR systems~\cite{fasoli21_interspeech} reported in the literature (line 9, 10) on the same task. The combination of the 4-bit TDNN-F acoustic and 2-bit Transformer language models (line 5) produced an overall “lossless” compression ratio of 13.6 over the full precision baseline (line 1) and comparable WER.
With the use of compact LHUC transforms of only 2000 parameters per speaker, speaker adaptation further reduced the WERs of the quantized TDNN-F systems by 1.5\%-2.7\% absolute on the NIST {\bf Hub5’00} test data (line 7-8 vs. 5-6). All our quantized TDNN-F systems (line 5-8) outperformed the \emph{IBM-BLSTM} system using the 300-hr SWBD data (line 9). The WERs of our \emph{TDNN-8bit+LHUC} system (line 8) are comparable to those of the 2000-hr SWBD+Fisher data trained \emph{IBM-BLSTM} system (line 10).

\vspace{-0.5em}
\section{Conclusions} 
This paper presents the development of a high performance and low-footprint 4-bit quantized LF-MMI TDNN based ASR system on the 300-hr Switchboard corpus. Fine-grained performance sensitivity at different model components to quantization errors was accounted for using a combination of neural architectural compression and mixed precision quantization approaches, which were also used to produce low-bit quantized Transformer LMs. An overall “lossless” compression ratio of 13.6 was obtained over the baseline full precision system while incurring no statistically significant WER increase. Future researches focus on improving hardware implementation and tighter integration with speech enhancement front-ends. 

\vspace{-0.5em}
\section{Acknowledgement}
This research is supported by Hong Kong RGC GRF grant No. 14200021, 14200218, 14200220, Innovation \& Technology Fund grant No. ITS/254/19 and ITS/218/21.

\bibliographystyle{IEEEtran}

\bibliography{mybib}


\end{document}